\documentclass[epsf,usegraphicx]{mn2e}

\usepackage{xspace}
\usepackage{amsfonts}
\usepackage{pifont}
\usepackage{hyperref}

\title[I band variability in ultra cool dwarfs] {Searching for $I$ band
  variability in stars in the M/L spectral transition region}

\author[]
{Gavin Ramsay$^{1}$, Pasi Hakala$^{2}$, J. Gerry Doyle$^{1}$ \\
$^{1}$Armagh Observatory, College Hill, Armagh, BT61 9DG\\
$^{2}$Finnish Centre for Astronomy with ESO (FINCA), University of Turku,
V\"{a}is\"{a}l\"{a}ntie 20, FI-21500 PIIKKI\"{O}, Finland\\
}

\date{Accepted 2015 July 28.  Received 2015 July 20; in original form 2015 May 20}

\begin{document}
\outer\def\gtae {$\buildrel {\lower3pt\hbox{$>$}} \over 
{\lower2pt\hbox{$\sim$}} $}
\outer\def\ltae {$\buildrel {\lower3pt\hbox{$<$}} \over 
{\lower2pt\hbox{$\sim$}} $}
\newcommand{\ergscm} {ergs s$^{-1}$ cm$^{-2}$}
\newcommand{\ergss} {ergs s$^{-1}$}
\newcommand{\ergsd} {ergs s$^{-1}$ $d^{2}_{100}$}
\newcommand{\pcmsq} {cm$^{-2}$}
\newcommand{\ros} {\sl ROSAT}
\newcommand{\chan} {\sl Chandra}
\newcommand{\xmm} {\sl XMM-Newton}
\newcommand{\kep} {\sl Kepler}
\def\rchi{{${\chi}_{\nu}^{2}$}}
\newcommand{\Msun} {$M_{\odot}$}
\newcommand{\Mwd} {$M_{wd}$}
\newcommand{\Lsol} {$L_{\odot}$}
\def\Mdot{\hbox{$\dot M$}}
\def\mdot{\hbox{$\dot m$}}
\newcommand{\teff}{\ensuremath{T_{\mathrm{eff}}}\xspace}
\newcommand{\tickYes}{\checkmark}
\newcommand{\tickNo}{\hspace{1pt}\ding{55}}

\maketitle

\begin{abstract}
We report on $I$ band photometric observations of 21 stars with
spectral types between M8 and L4 made using the Isaac Newton
Telescope. The total amount of time for observations which had a
cadence of $<$2.3 mins was 58.5 hrs, with additional data with lower
cadence. We test for photometric variability using the Kruskal-Wallis
H-test and find that 4 sources (2MASS J10224821+5825453, 2MASS
J07464256+2000321, 2MASS J16262034+3925190 and 2MASS
J12464678+4027150) were found to be significantly variable at least on
one epoch. Three of these sources are reported as photometrically
variable for the first time. If we include sources which were deemed
marginally variable, the number of variable sources is 6 (29
percent). No flares were detected from any source. The percentage of
sources which we found were variable is similar to previous studies.
We summarise the mechanisms which have been put forward to explain the
light curves of brown dwarfs.
\end{abstract}


\section{Introduction}

{\sl Ultra Cool Dwarfs} (UCDs) are generally defined as having a
spectral type later than M8. The temperature of stars on the boundary
between M and L spectral classes is $\sim$2200 K and the corresponding
mass is $\sim$0.075 \Msun ($\sim$80 Jupiter masses) (e.g. Basri et al
2000).  This is the mass where a star cannot burn hydrogen in its core
and are known as brown dwarfs. The 2MASS survey led to the
  detection of increasing numbers of these objects (e.g. Kirkpatrick
  et al 1999), with SDSS (e.g. Geballe et al. 2002), DENIS
  (e.g. Kendall et al. 2004) and WISE (e.g. Kirkpatrick et al. 2011)
  all leading to many additional discoveries. There are now nearly 2000
  brown dwarfs which have a spectral type later than
  M9{\footnote{\url{http://www.johnstonsarchive.net/astro/browndwarflist.html}}.}

UCDs have provided a number of surprise's over the last few decades. A
small number of UCDs show narrow-band pulsed radio emission which is
100\% circularly polarised (e.g. Berger et al. 2001, Hallinan et
al. 2008, Doyle et al. 2010). These pulses are thought to be produced
by electron cyclotron maser emission and have an important diagnostic
power to determine the stellar magnetic field strength and
topology. The recent detection of a flare from a T6.5 dwarf (Route \&
Wolszczan 2012) implies a field strength of 1.7 kG which poses severe
challanges for dynamo theories for magnetic field generation.

Observations of UCDs also allow the study of the atmospheric chemistry
of very cool stars which has direct implications of our understanding
of exo-planet atmospheres. At temperatures lower than $\sim$2600
  K, clouds begin to form and their depth and composition is expected
  to change with temperature (see for example Tsuji 2002, Allard,
  Homeier \& Freytag 2012). However, there is still some controversy
  regarding the origin of periodic variations seen in some late type
  dwarf stars. Optical observations of TVLM 513-46546 (one of the
  radio pulsing systems which also shows X-ray emission) made over 7
  years showed a stable rotation period of 1.96 hrs (Doyle et
  al. 2010, Wolszczan \& Route 2014), which was taken to be evidence
  for a rotating magnetic spot. However, simultaneous $g$ and $r$ band
  data showed that their light curves were anti-correlated, which may
  point to long lived atmospheric dust clouds (Littlefair et
  al. 2008), although an auroral hot spot is an alternative
  explanation (Nichols et al. 2012).

Since the turn of this century there have been a number of photometric
variability surveys of UCDs in both optical and IR bands (see,
  for instance, Radigan et al 2014, Wilson, Rajan \& Patience 2014,
  Metchev et al. 2015).  The optical surveys have largely been
conducted in the $I$ band and are most suitable to observe stars
earlier than mid L spectral types, whilst IR observations have
focussed largely on later spectral types, since these stars become
very faint in the $I$ band.  Given these surveys have a wide range of
cadence and duration, definitive conclusions have been hard to
make. However, it is now becoming apparent that in IR bands $\sim$40
percent of sources with spectral types on the border between L and T
types show strong variability on the timescale of rotation period,
whilst $\sim$60 percent of L/T sources outside this region show
variability with an amplitude of $\sim$1 percent (Radigan et
  al. 2014). In the $I$ band, surveys show that many objects show low
level variability but rather few show evidence of rotational
signatures.

To increase the general body of knowledge of the photometric
variability of sources in the M/L spectral transition region and
search for sources which show a signature of a rotation period, we
conducted a survey of 21 sources (some of which have not been studied
before) using the 2.5 m Isaac Newton Telescope (INT) on the island of
La Palma in 2014 and 2015 to search for variability on the several
hour timescale. We present the results here and compare our findings
with previous studies.

\section{Observations}

\begin{table*}
\caption{The sources which we observed using the INT, where the source
  2MASS identification is given in the first column. The i mag is
  taken from the SDSS and for others we infer the approximate $I$ band
  mag (given in italics) from their 2MASS colours and spectral type
  (Liebert \& Gizis 2006). We also indicate the spectral type where
  have searched the dwarfarchives.org database and supplemented this
  with information from Liebert \& Gizis (2006), Jenkins et
  al. (2009), West et al. (2011) and Zhang et al. (2010). The start of
  our INT observations is given in MJD, the duration of these
  observations and the number of photometric points. These are
  variable since in the earlier observations we alternated between a
  number of other filters. The INT Variability Column indicates
  whether the light curve is significantly variable (Y), not variable
  (N) or marginally variable (M).}
\begin{center}
\begin{tabular}{llllllr}
\hline
Object & Mag & Spectral & Observation Start & Duration & No. Pts & INT \\
       & (i/$I$) & Type & (MJD)             & (mins)   &         & Variability?\\
\hline
2MASS J02281101+2537380 & 18.2       & L0    &  56931.957 & 160 & 21 & N\\
2MASS J02511490--0352459& {\sl 16.4} & L1    &  56932.091 & 163 & 22 & N \\
2MASS J02522628+0056223 & 17.1       & M9    &  56932.970 & 187 & 25 & N \\
                        &            &       &  56934.102 & 205 & 188 & N \\
2MASS J03202839--0446358& {\sl 16.6} & L0.5  &  56933.108 & 140 & 16 & N\\
2MASS J19064801+4011089 & 18.2       & L1    & 56933.832 & 154 & 139 & N \\
                        &            &       &   56935.938 & 130 & 120 & N\\
2MASS J00154476+3516026 & 14.8       & L2    & 56933.974 & 165 & 151 & N \\
2MASS J19165762+0509021 (VB10) & {\sl 12.6} &  M8     &  56934.841 & 164 & 151 & N\\
2MASS J01353586+1205216 & 18.8       &  L1.5          &  56935.077 & 132 & 121 & N\\
2MASS J03205965+1854233 (LP 412-31) & 15.7 & M8       &  56935.174 & 107 & 99 & N \\
2MASS J18071593+5015316 & {\sl 16.2} & L1             &  56935.836 & 130 & 120 & N\\
2MASS J03395284+2457273 & {\sl 15.5} & M8   &   56936.078 & 120 & 110 & N\\
2MASS J03521086+0210479 & {\sl 16.4} & L0   & 56936.172 & 120 & 110 & N\\
2MASS J10224821+5825453	& 17.9       & L1   & 57112.878 & 209 & 142 & Y\\
                        &            &      & 57115.984 & 230 & 121 & Y\\
2MASS J13004255+1912354 (MN Com ) & 17.2 & L3   &  57113.131 & 165 & 116 & M\\
2MASS J07464256+2000321 & 16.1       & L0+L1.5       & 57113.927 & 98 & 80 & N\\
                        &            &          & 57115.850 & 180 & 141 & Y\\
2MASS J07472762+2608406 & 18.7       & M9.5     & 57114.852 & 130  & 58 & N\\
2MASS J11083081+6830169 & {\sl 16.4} & L1       & 57114.946 & 270 & 121 & M\\
2MASS J16262034+3925190 & 17.9       & L4       & 57115.141 & 154 & 70 & N \\
                        &            &          & 57116.154 & 135 & 72 & Y\\
2MASS J08175266+1947279 & 19.5       & M9       & 57116.849 & 153 & 67 & N\\
2MASS J12464678+4027150 & 19.8       & L4       & 57116.975 & 166 & 75 & Y\\
2MASS J13364062+3743230 & 18.7       & L1       & 57117.113 & 191 & 84 & N\\
\hline
\end{tabular}
\end{center}
\label{sources}
\end{table*}

In selecting targets we used the {\tt SIMBAD} database to obtain an
initial broad target list of sources with spectral type in the range
M8--L4. Since we had time on the INT at two different epochs we made a
smaller target list for each epoch and then ensured that the observed
sources were as distant as possible from the moon on the given date
(the moon was close to full at each epoch). Some sources were chosen
so that previous variability information was available for comparison
and some sources had no previously published photometry. The details
of the sources observed are shown in Table \ref{sources}.

Observations were made using the INT with the Wide Field Camera and
the Sloan Gunn $i$ band filter between 2014 Oct 1--5 and 2015 Mar 31
-- Apr 4. To reduce the readout time of the detector we only used a
sub-array of Chip4, giving a typical readout time of 14 sec. The
exposure time depended on the source brightness but was typically 2
min. (In Table \ref{sources} we note the $i$ or $I$ band mag of each
source). For the first four targets we also obtained data using
$grH\alpha$ filters which resulted in a cadence of $\sim$8 mins
(Given the sources are very faint in these observations and
  the cadence was poor they do not provide additional information).
For the rest of the observations the cadence was 1.1--2.3 mins. The
total amount of $i$ band data was 4158 min (=69.3 hrs) -- the amount
of time where the cadence was $<$2.3 min was 3508 min (=58.5 hrs).

The images were reduced using standard {\tt FIGARO} (Shortridge et al
2004) routines. We obtained twilight sky flat fields and additional
observations were made during the night of blank fields which allowed
the removal of the effects of fringing. (On some nights thin drifting
cloud was present which prevented the complete removal of fringing
effects in each science image).

\begin{figure*}
\begin{center}
\setlength{\unitlength}{1cm}
\begin{picture}(6,11.6)
\put(15,-0.6){\includegraphics{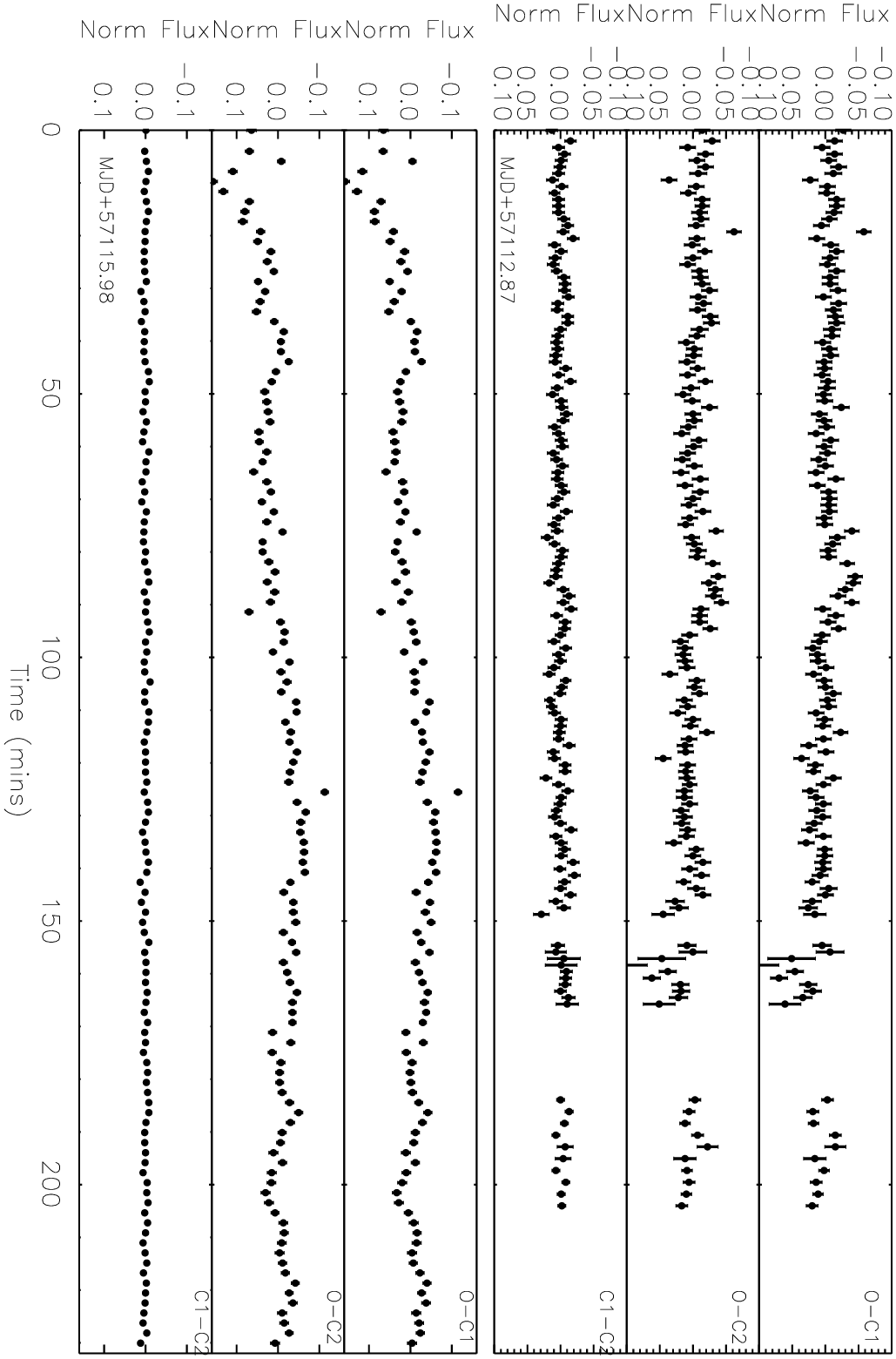}}
\end{picture}
\end{center}
\caption{The light curve of 2MASS J1022+5825 made using the INT in the
  $I$ band taken on two different nights in April 2015.}
\label{light} 
\end{figure*}

For each set of observations we performed aperture photometry of the
target and two comparison stars in the field using the {\tt PHOTOM}
package (Eaton, Draper \& Allen 2009) which is part of the {\tt
  STARLINK}\footnote{\url{http://starlink.eao.hawaii.edu/starlink}} suite of
software. Given the very red intrinsic colour of our targets, it was
likely that these comparison stars would have different colours
compared to the target. In some fields there were low numbers of
potential comparison stars. However, for all observations we obtained
photometry of the target and two comparison stars and derived three
differential light curves: O-C1, O-C2 and C1-C2 (where O is the object
and C1 and C2 is the first and second companion star). As an example
we show the light curve for 2MASS J1022+5825 in Figure \ref{light}
made on two nights in April 2015, (this source was subsequently
  found to be variable at both epochs).

\section{Testing for Variability}

One of the standard tests for variability in photometric light curves
is based on evaluating the null hypothesis of `no variability' by
least squares fitting the light curve with a zeroth order polynomial
(i.e. determine the best fitting constant level). One can then compute
how well the constant model agrees with the data (e.g. the $\chi^{2}$
test). Whilst this is a very simple and straight forward method, it
suffers from severe problems such as flagging all light curves as
variable if there are systematic errors and/or outliers (non-Gaussian
data) involved. It is therefore beneficial to employ a method that is
less sensitive to these effects.

We have chosen to employ the non-parametric Kruskal-Wallis H-test
(Kruskal \& Wallis 1952) to search for variability in our light
curves. The H-test divides the light curve into N different samples
(e.g. time windows) and after ranking the light curve values by flux,
the test then measures the distribution of flux ranks in different
samples. We defined time windows so that there were on average ten
photometric points per window. This allows us to decide whether the
population means in different samples are identical without assuming
them to follow a Gaussian distribution. Furthermore, the errors of
individual points are not used. These assumptions help to overcome
some of the systematics, although clear systematic trends in the light
curves will still produce false detections.

In order to minimize the number of false detections, we have employed
two comparison stars for each target star (Finding charts for all
targets can be accessed in the Supplementary material where comparison
stars are indicated -- see the end of this section for how to obtain
this material). This way we can compute the H-test on differential
magnitudes of O--C1, O--C2 and C1--C2. We can then compare the
p-values of those three tests to filter out false variables. For
instance, if both O--C1 and O--C2 tests give very low value for the
null hypothesis of `no variability', but also C1--C2 gives a very low
value, any variability is most likely due to a systematic
effect. However, on the other hand, if the C1--C2 p-value is much
larger than the O--C1 and O--C2 ones, we are likely to have a real
variable. We first identified those sources which had differential
light curves O--C1 and O--C2 which were constant with a probability of
$<1\times10^{-3}$ (i.e. they were likely to be variable). Those
sources then had to have a probability that the C1--C2 differential
light curve was constant with a much higher probability. We then next
visually inspected these light curves to determine if they showed any
trends caused by differential colour extinction -- none did.  We
indicate in Table 1 whether a source is classed as `variable',
`non-variable', or `marginally' variable.  (The light curves and all
cleaned CCD images can be accessed from information given in the
Supplementary material which will be available on the MNRAS website
and also \url{http://www.arm.ac.uk/~gar/intdwarf}).

\begin{figure*}
\begin{center}
\setlength{\unitlength}{1cm}
\begin{picture}(10,8)
\put(16,-4){\includegraphics{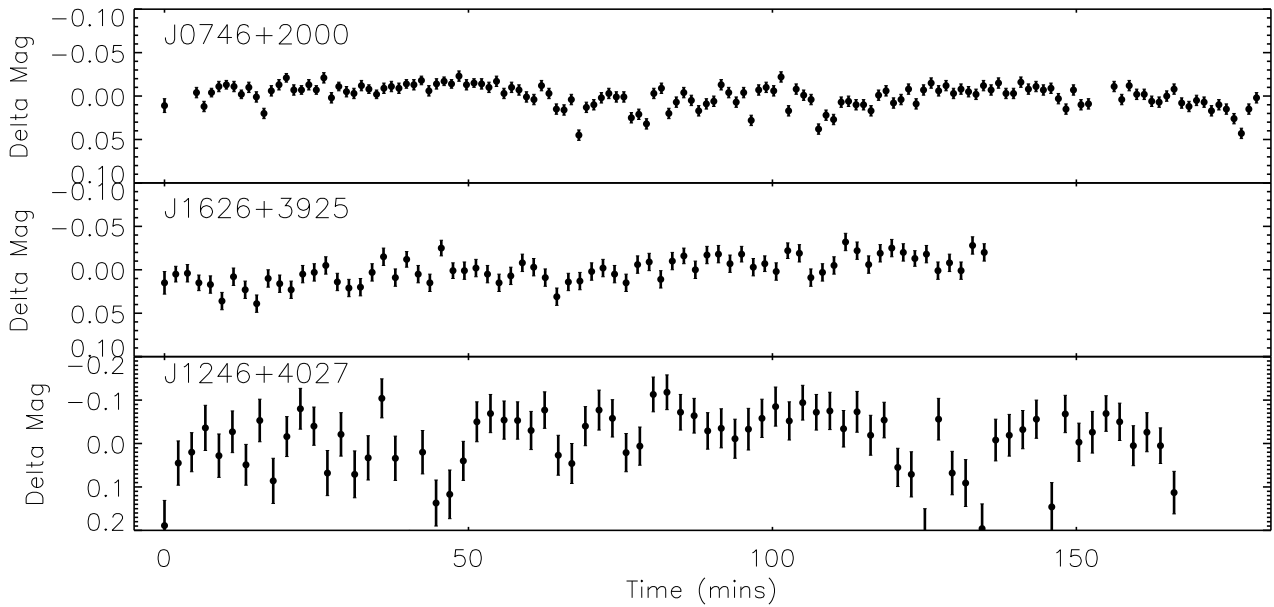}}
\end{picture}
\end{center}
\caption{The light curve of the three sources, J0746+2000,
    J1626+3925, J1246+4027, which in addition to J1022+5825, were
    found to be photometrically variable.}
\label{light2} 
\end{figure*}

\section{Results}

Out of the 21 M/L stars which we observed using the INT, 4 (19
percent) show variability at one epoch. The Kruskal-Wallis H-test
indicates they are constant with a probability of $<2\times10^{-4}$.
There are two additional sources which were also initially classed as
variable but their C1-C2 differential light curve showed evidence for
variability and were therefore classed as `marginally' variable. If we
include these marginal sources the number of variable sources in our
sample becomes 6 sources (or 29 percent).  Our small sample precludes
us from determining whether sources of a given spectral type were more
likely to be variable or not. The fraction of sources which we find to
show photometric variability is very similar to that found by Koen
(2013) in a larger sample who found that 26 percent of M stars and 23
percent of L stars showing $I_{c}$ band variability over the same
timescale.

Of the four sources which were classed as variable, one (J1022) was
variable at each of the two epochs (Figure \ref{light}) it was
observed while J0746 and J1626 were found to be variable for only one
of the two epochs they were observed (Table \ref{sources}). (For those
sources which were observed at two epochs, none showed significant
differences in their mean flux).  (We show the light curves for
J0746+2000, J1626+3925 and J1246+4027 in Figure \ref{light2}). None of
these `variable' light curves show evidence of periodic behaviour.
The only source which shows short duration trends (J1246) is faint
($I\sim$19.8) and would be a good candidate for further longer
duration observations.

The origin of the variability of late M/early L type stars is
  still a matter of some debate, with starspots, holes (or `spots') in
  in the stellar atmosphere (of the sort seen in the Jovian planets)
  or inhomogeneous cloud layers all being proposed.  As the star
  rotates these features can cause photometric
  variations. Disentangling the effects is difficult but identifying
  sources which do show variability provide opportunities to make more
  detailed investigation of the causes. The fact that we find only
  $\sim$1/4 of our targets to be photometrically variable suggests
  that most stars on the M/L boundary have fairly homogenous
  clouds. Stars which {\it do} show variability are more likely to
  have inhomogenous cloud layers or spots. In the Spitzer survey of 44
  L3-T8 dwarfs, Metchev et al. (2015) conclude that the reason for the
  large number of irregular variables in their sample was due to rapid
  changes in the distribution of spots. This may also be the case here
  although at least some of our targets are known binaries. We now
  briefly comment on the sources in our survey which do show
  variability.

Although J1022 has shown variability in the profile of the H$\alpha$
emission line (Reiners \& Basri 2008) it did not show long term $K$
band variations in the study of Lopez Marti \& Zapatero Osorio (2014).
J1022 appears to be older than $\sim$500 Myr and has a mass close to
above the Lithium burning limit (Zapatero Osorio et al 2014). This
appears to be the first detection of photometric variability on the
several hour timescale.

J0746 shows strongly polarised radio emission on a period of 2.07 hrs
(Berger et al 2009) and its optical light curve shows a period of 3.3
hrs (Harding et al 2013). Since J0746 is a known binary system
(L0+L1.5) it is thought that J0746A is the source of the 3.3 hr
modulation, while J0746B is the source of the radio pulses. Our first
set of observations only covered 98 mins and no significant
variability was detected but in the second set of observations we
detect a significant modulation (based on the false alarm probability)
on a period of 1.65 hrs -- this is half the rotation period reported
in Harding et al (2013). Given the observation duration was only 180
mins it is not surprising we do not identify the true rotational
period. The other two variable sources, J1626 and J1246, do not appear
to have been targeted for photometric variability before, and J1626
was found to be variable on one of the two epochs.

{\sl Kepler} observations of the L1 dwarf WISEP J190648.47+401106.8
showed white light flares with energies in the range 6$\times10^{29}$
-- 1.6$\times10^{32}$ erg and durations 3--160 min (Gizis et al
2013). We would have expected to easily detect a flare with an
amplitude of 0.05 mag in our data. Assuming a luminosity of
1.4$\times10^{28}$ erg/s in the {\sl Kepler} pass-band for an L1 star
(Gizis et al 2013) and a flare lasting 10 min and a peak amplitude of
0.05 mag, this would imply a luminosity $\sim3\times10^{29}$
erg. Gizis et al (2013) detected a flare with $E\sim8\times10^{29}$
erg every $\sim$100 hrs. Given our high cadence data covered a total
duration of nearly 60 hrs the fact we did not detect any flares is not
surprising.

\section{Conclusions}

Our survey of $I$ band photometric variability in 21 stars with
spectral types covering the M/L transition region show that 19 percent
show clear variability over the several hour timescale. If we include
sources which show marginal evidence of variability this number
increases to 29 percent. This is very similar to the study of Koen
(2013) who found 26 percent of M stars and 23 percent of L stars
showing $I_{c}$ band variability. Three of our variable sources have
not been shown to be photometrically variable before. This makes them
prime candidates for radio surveys searching for highly polarised
radio emission in UCDs.

It is clear that observations covering a longer timescale are
essential to obtain more robust information on the variability of M
and L dwarfs on different timescales and (perhaps most importantly)
identify periodic behaviour. One such facility is the Next Generation
Transit Survey (NGTS, Wheatley et al 2013) which is a wide field
survey with a main goal of detecting sub-Neptune sized exo-planets,
but given its red sensitivity it will be very well suited to long term
monitoring of M/L dwarfs. Such facilities are not only better suited
to searching for variability and will help inform models (e.g. Cooper
et al. 2003, Helling et al. 2008, Stark, Helling, Diver 2015).

\section{Acknowledgements}

We thank the anonymous referee for a helpful and constructive report.
The paper is based on observations made with the Isaac Newton
Telescope operated on the island of La Palma by the Isaac Newton Group
in the Spanish Observatorio del Roque de los Muchachos of the
Instituto de Astrofísica de Canarias. This research has made use of
the SIMBAD database, operated at CDS, Strasbourg, France. Armagh
Observatory is supported by the N. Ireland Government through the Dept
Culture, Arts and Leisure.

\end{document}